\documentclass[12pt,preprint]{aastex}

\usepackage{ulem}
\usepackage{color}

\newcommand{\water}{\rm{H_2O}}
\newcommand{\methane}{\rm{CH_4}}
\newcommand{\methanol}{\rm{CH_{3}OH}}
\newcommand{\formaldehyde}{\rm{H_{2}CO}}
\newcommand{\vinylcyanide}{\rm{CH_{3}CN}}
\newcommand{\methrad}{\rm{CH_3}}
\newcommand{\molh}{\rm{H_2}}

\slugcomment{To appear in ApJ.}

\shorttitle{Lukewarm Corino Models of L1527}
\shortauthors{Hassel, Herbst, \& Garrod}

\begin{document}

\title{Modeling the Lukewarm Corino Phase - is L1527 unique?}

\author{George E. Hassel}
\affil{Department of Physics, The Ohio State University,
    Columbus, OH 43210}
    \email{ghassel@mps.ohio-state.edu}
\author{Eric Herbst}
\affil{Departments of Physics,  Astronomy,  and Chemistry, The Ohio State University,
    Columbus, OH 43210}
\and
\author{Robin T. Garrod}
\affil{Max-Planck-Institut f\"{u}r Radioastronomie, Auf dem H\"{u}gel 69, 53121 Bonn, Germany}

\begin{abstract}
Sakai et al.~have observed long-chain unsaturated hydrocarbons and cyanopolyynes in the low-mass star-forming region L1527, and have attributed this result to a gas-phase ion-molecule chemistry, termed ``Warm Carbon Chain Chemistry'',   which occurs during and after the evaporation of methane from warming grains.
 The source L1527 is an envelope surrounding a Class 0/I protostar with regions that possess a slightly elevated temperature of $\approx$~30 K.  The molecules detected by Sakai et al.~are 
typically associated only with dark molecular clouds, and not with the more evolved hot corino phase.  In order to determine if L1527 
is chemically distinct from a dark cloud, we compute models including various degrees of heating.  The results indicate that the composition of 
L1527 is somewhat more likely to be due to ``Warm Carbon Chain Chemistry'' than to be a remnant of a colder phase.  If so, the molecular products
provide a  signature of a previously uncharacterized early phase of low mass star formation, which can be characterized as a ``lukewarm''  corino.   We also include
 predictions for other molecular species that might be observed toward candidate lukewarm corino sources.  Although our calculations show that unsaturated hydrocarbons and cyanopolyynes can be produced in the gas phase as the grains warm up to 30 K, they also show that such species do not disappear rapidly from the gas as the temperature reaches 200 K, implying that such species might be detected in hot corinos and hot cores.
\end{abstract}

\section{Introduction}

Recent observations of the warm gas associated with L1527 by \citet{sea07}(hereafter SSHY)~have revealed an unexpected chemical composition including unsaturated chain radicals of formula $\rm{C_{n}H}$,~carbenes of formula $\rm{C_{n}H_{2}}$, and cyanopolyynes, $\rm{HC_{2n}CN}$.  These molecules are typically associated with the cold, dark cloud stage of star formation \citep{suz92,mmc00,hmy04,shc04}, but independent observations indicate that the physical state of L1527 is more dense and slightly warmer than dark clouds.  Indeed, the column densities of 13 species detected by SSHY toward L1527 are only slightly lower than those detected toward the dark cloud TMC-1.  The presence of such unsaturated species indicates that L1527 may be one example of a chemically unexplored, transitional phase of low-mass star formation.  The analogy with dark clouds is strengthened by the detection of the anions C$_6$H$^-$ \citep{ssoy07} and, most recently, C$_4$H$^-$ \citep{ssy07,agund08}.

The physical structure of L1527 contains  an envelope surrounding the low-mass protostar IRAS 04368+2557 with the object
between  the Class 0 and I protostellar phases \citep{awb00}.  According to \citet{mm98},  the envelope is undergoing inside-out collapse in good agreement with the \citet{shu77}~model.  The source also shows evidence for bipolar outflows in the plane of the sky \citep{ohhm97}.  The SSHY observations include a description of a profile mapping technique, indicating that the velocity of $\rm{C_{4}H}$~is consistent with the central protostar rather than the outflow region. This region differs from previously investigated star-formation phases because the envelope is only slightly elevated in temperature.   Based on a fit to central peak emission and the SED profile, \citet{hs00} determined a  temperature of 35 K, while \citet{lrdwh02} obtained a temperature of 22 K via a similar computation for the central region of L1527. These temperatures are slightly warmer than a typical 10 K cold cloud but colder than a hot core/corino region.  Selected temperatures in these evolved regions have been estimated as $T= 50-80$~K for the low mass hot corinos in NGC1333 \citep{mea04}, $T > 100$~K for the IRAS 16293 hot corino \citep{ccchlmt00}, and $T=150-200$~K for the G327.3-0.6 high mass hot core \citep{gniwb00}.  In addition to observations of its physical state, L1527 was included among 18 dense cores in the molecular survey of \citet{jsv04} (hereafter JSV04).  These authors detected the molecular species CO, CN, CS, HCN, HNC, HC$_3$N, SO, HCO$^+$, and N$_2$H$^+$ towards L1527.  Although this set of observations does not include hydrocarbons or complex cyanopolyynes and does not include velocity variations, it does provide an independent basis for model comparison. It is clear that further interferometric study of L1527 will provide much more information toward understanding the nature of this source.   

While the presence of non-saturated hydrocarbons is more typically a dark cloud characteristic, the hot core/corino evolutionary phases are more commonly associated with more saturated, oxygen- and nitrogen-rich species.  What can explain the dark-cloud species detected in L1527?  One possibility is that these molecules are merely remnants of the colder pre-stellar core phase that survive a very fast collapse towards the center of the object.  In the dense portions of such sources, however,  it is expected that the carbon chain molecules would freeze out to grain surfaces more rapidly than in a cloud of lower density.  As noted in SSHY,  the slight elevation in temperature is capable of  regenerating abundances of carbon chain molecules after they have frozen out to grain surfaces.  The mechanism for this process is more likely the selective release and subsequent reaction of highly volatile molecules such as methane (CH$_4$) from grain surfaces than the actual sublimation of carbon chains formed by cold processes.  For this reason, the process was named ``Warm Carbon Chain Chemistry'', or WCCC (SSHY).    

In this paper, we present detailed models of a low-mass star-forming region with a slightly elevated temperature, for which we use the phrase ``luke-warm corino'', where the adjective ``luke-warm'' was previously used in talks by E. van Dishoeck.  Specifically, we attempt to determine if the observations can be explained by models of a cold region alone, or if the chemistry is the signature of a newly discovered protostellar phase that requires only a small amount of heating.    In particular, we consider  models in which a cold dense phase is followed by a warming to temperatures of 30 K and beyond, based both on the dust warming and on molecular line observations (SSHY).  The method used is similar to that discussed by \citet{gh06} for hot cores and corinos, in which a gas-grain chemical model is utilized and both surface and gas-phase reactions are considered during the cold and heat-up phases.   If L1527 is a representative example of a simple, isolated, collapsing core, perhaps more low-mass protostellar sources should exhibit a similar lukewarm envelope and the WCCC chemical signature.

\section{Model}
\label{modelsection}

We utilize the OSU gas-grain code, first described by \citet{hhl92}, with some more recent modifications, the most important of which is the capability to increase the gas and dust temperatures with time \citep{gh06}.  The  model includes the granular parameters listed in Table~\ref{tbl-gr}.  We include  the Rice-Ramsperger-Kessel (RRK) approach for reactive desorption from the solid phase, first suggested by \citet{gpch06}.  Reactive desorption, or non-thermal desorption of the products of exothermic surface reactions, is regulated by the parameter $a_{\rm{RRK}}$, approximately equal to the efficiency of the process.  For the models reported here $a_{\rm{RRK}}=0.01$, as chosen by \citet{gwh07}.  The results of corresponding models with $a_{\rm{RRK}}=0$~are not significantly different from those presented, and are thus omitted for simplicity.   In addition to the previously published modifications of the gas-grain network, we added the photodesorption of CO ice, based on the laboratory measurements of  \citet{ofafsvl07}. The desorption can be caused by both the ambient UV field and by secondary photons generated by cosmic rays.    Because L1527 is a dense region, and not UV-dominated, we assume a typical interstellar radiation field and a value $A_{\rm{v}}=10$ for all models presented.  Consequently, the inclusion of CO photodesorption does not introduce a significant departure from results in which it is not included.  The network includes 6312 reactions involving a total of 655 gaseous and surface species.  Negative molecular ions have not yet been included in the model; an explanation of their abundances in L1527 is being undertaken using a purely gas-phase model.

Our overall method is to follow the chemical evolution of a homogeneous (one-point) parcel of material as it progresses through the cold pre-stellar core phase and then heats up. It is assumed that the dust and gas temperatures are in equilibrium with one another. This approach differs from the hydrodynamic multi-point collapse model of \citet{aikawa}~in that we make simplified estimates of the time dependence and final value of the temperature  and focus on the details of chemical evolution of groups of molecules during both the cold and warming-up phases.  The advantage of our approach is that we are not wedded to a particular dynamical time dependence, the disadvantage is that it is unclear whether the physical changes that occur are totally realistic.  

\begin{figure}
\includegraphics[angle=90,scale= 0.8]{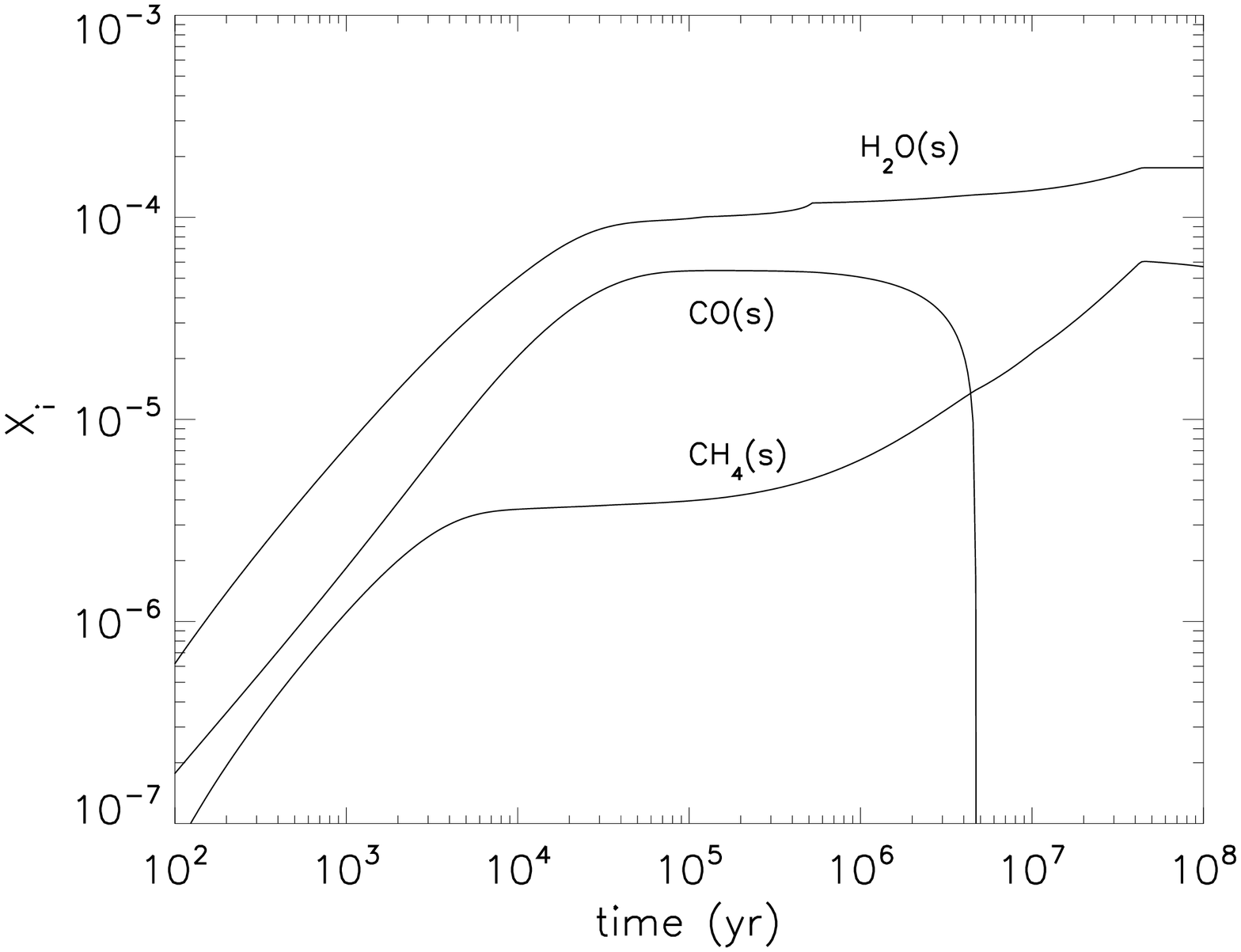}
\caption{Surface fractional abundances of major mantle species with respect to total hydrogen at $n_{\rm{H}}=10^6~\rm{cm}^{-3}$ and $T$ = 10 K.  \label{fig-init}}
\end{figure}

We start the calculations by simulating the gas-grain chemistry that occurs in the central region of a cold pre-stellar core with a constant density of $n_{\rm{H}}=10^{6}~\rm{cm}^{-3}$ and an initial temperature of $T=10$~K.  We start with gaseous atoms and ions except for molecular hydrogen, with abundances following low-metal elemental abundances as listed in Table~\ref{tbl-ab} \citep{gwh07}.  In addition, we consider the effect of an increased sulfur abundance of $1\times10^{-5}$, initially in the form of S$^+$, and we discuss the results of this variation below.   Figure~\ref{fig-init} shows the fractional abundances of the major ice species H$_2$O(s), CO(s), and CH$_4$(s) as functions of time under  these conditions.   The surface abundance of the predominant component, $\water$(s), reaches a steady-state fractional abundance of $\approx$ 10$^{-4}$  after a little more than 10$^{4}$ yr, while both  $\methane$(s) and CO(s) reach a quasi-steady state at this time.  After 10$^{6}$ yr, CO(s) decreases dramatically because it is hydrogenated, while $\methane$(s) actually increases in abundance.  The initial cold phase in our warm-up models is allowed to continue for $10^{5}$~yr, corresponding to the maximum CO(s) abundance, but we also run models to $10^{4}$~and~$10^{6}$~yr for comparison.  The abundance of $\methane$(s) in this range with respect to $\water$(s) is calculated to be $\approx$ 3.5\%, which is in agreement with recent $Spitzer~Space~Telescope$ observations according to SSHY, and with previous $ISO$ results of 1-4\% reported by \citet{gibb04}.  More recently, \citet{oberg08} reported that the CH$_4$(s) abundance may vary by 2-13\% among a variety of sources, but that it is more narrowly defined to be $4.7 \pm 1.6\%$~if $N_{\rm{H_{2}O(s)}} > 2\times 10^{18}$~cm$^{-2}$.  Our calculation is in strong agreement with this new result.  This extent of variation does not strongly affect the resulting abundances of the carbon-chain molecules to the best of our estimation.  But, we cannot easily vary the amount of solid methane in our  gas-grain model, since it is determined by the chemical history of the cold stage.  In a gas-phase model in preparation (Harada \& Herbst), we are looking into the dependence of assorted abundances on the gas-phase abundance of methane immediately after evaporation.

After the initial cold phase, the gas and dust temperatures are increased using the expression \citep{gh06}
\begin{equation}
T=T_{0}+(T_{\rm{max}}-T_{0})\left({{\Delta t} \over {t_{\rm{h}}}}\right)^{n},
\end{equation}
where $T_{0}$ is the initial temperature of 10 K, $T_{\rm{max}}$ is the maximum temperature, $t_{\rm{h}}$ is the heating timescale, and $n$ is the order of the heating.  If L1527 is in a transitional evolutionary stage between cold and hot phases, the temperature may be increasing above 30 K.  To account for this, the simulated temperature increase is carried to three maximum temperature values of $T_{\rm{max}}=30$~K, 100 K, or 200 K in separate model runs.  After reaching the asymptotic maximum, the model holds the temperature at that constant value while continuing to trace the chemical evolution.  The adopted heating timescale is $2 \times 10^5$~yr with an order of $n=2$ for each $T_{\rm{max}}$, based on the estimated warm-up timescale for intermediate star formation.  \citet{gh06}~primarily utilized this timescale to simulate the chemistry during the formation of hot cores. 

\section{Results}
Our results are obtained as fractional abundances with respect to $n_{\rm H}$, $X_{i}=n_i/n_{\rm{H}}$.   The SSHY observations were reported as column densities, so comparison with observation requires us to convert these using the total hydrogen column density $N_{\rm{H}}=6\times 10^{22}$~cm$^{-2}$ \citep{jsv02}.  The converted observational values are listed in Table~\ref{tbl-peak} along with the observed fractional abundances from the earlier study of \citet{jsv04} and presented where appropriate in figures discussed below.  The results of our simulations can be plotted as temporal evolutions of molecular abundances, both in the gas and on granular surfaces,  both during the cold initial period and the subsequent warm-up, if it occurs.   Two representative examples for unsaturated carbon chain gaseous species are shown in Figure~\ref{fig-c4hex}, for C$_4$H, and Figure~\ref{fig-hc5nex}, for HC$_5$N.  In these figures, we display the results for constant temperature, $T=10$~K, and for heating to $T_{\rm{max}}=30$~and 200 K.  Models with similar heating but without surface reactions are also shown, but they do not show any significant distinction among temperature variations.   These example plots also include the SSHY observations as solid horizontal lines and the temperature profiles of the warm-up models.  The heating onset time was chosen to be 10$^{5}$ yr, as discussed above.  Similar procedures that varied the heating onset time between $10^{4}$~and $10^{6}$~yr showed no substantial difference in the results as a function of time after the start of heat-up.

The dominant features for C$_{4}$H and HC$_{5}$N in models in which heating begins to occur after 10$^{5}$ yr include (i) a ``cold'' peak at 10$^{4-5}$ yr, which is essentially the ``early-time'' peak of cold gas-phase models, followed by a decrease due both to gas-phase reactions and accretion onto dust particles, (ii) an increase in abundance that occurs when volatile molecules such methane and carbon monoxide begin to sublimate from the grain surfaces as the temperatures nears 30 K, and (iii) a second, or warm, peak, possibly with fine structure, which occurs in the time frame 10$^{5-6}$ yr, and (iv) a final decrease in abundance.  The second peak occurs at or near the period when the maximum temperature in the model (30 K or 200 K) is reached.  The warm peak abundance for C$_{4}$H and HC$_{5}$N is larger for the higher temperature model; in addition, the abundances when the higher temperature model is temporarily at 30 K are very similar to the abundances achieved when the 30 K warm-up model reaches its asymptotic temperature.  In the case where $T_{\rm{max}}=200$~K,  a satellite warm-up peak occurs near $T=30$~K and has abundance similar to that of the single $T_{\rm{max}}=30$~K warm-up peak.  The second, larger peak occurs when the temperature first reaches 200 K and the majority of the ice mantle species are driven from the grains.

In the cold model, the abundances of C$_{4}$H and HC$_{5}$N are slightly regenerated in a plateau region beginning at $t\approx 6\times10^{5}$~yr until $t\approx 4\times10^{6}$~yr as a result of reactive desorption.
Actually, the regenerated abundances of most molecules do not exceed the initial cold peak with the exceptions of $\rm{C_{10}}$, $\rm{C_{5}N}$, CS, $\rm{H_{2}CS}$, and $\rm{H_{2}S}$.  The species CN and HCO$^+$~do not decrease from the initial cold peak but remain at a plateau value for nearly $8\times 10^{6}$~yr.  Finally, as exemplified by C$_{4}$H and HC$_{5}$N, the warm up in the absence of surface chemistry shows no warm peak, demonstrating that surface chemistry prior to warm-up is indeed necessary for the production of gas-phase molecules in the vicinity of 30 K.  The detailed mechanism for the production of the gas-phase molecules during the warm-up is discussed below.

 In Table \ref{tbl-peak} we list peak abundances for a large number of species for four models in addition to observed fractional abundances for L1527 if available.   There are three columns that specify a maximum asymptotic temperature of 30 K, 100 K, and 200 K.  Here the peak values refer always to that achieved during warm up after $t=10^{5}$~yr and not in the cold phase.  In models with an asymptotic temperature that exceeds 30 K, the peak values can be greater than values in the vicinity of 30 K, as is the case for most unsaturated carbon chain species (e.g., C$_{4}$H and HC$_{5}$N)  as well as all standard hot-core-type molecules.  The potentially controversial high-temperature results for the unsaturated species are discussed in Section 3.3.  In the column in which results from a model with a constant temperature of 10 K are listed, the peak abundances are  the cold, ``early-time'' values prior to $t=10^{5}$~yr.  One can see that for unsaturated classes of organic molecules such as the radicals C$_{\rm n}$H, and the cyanopolyynes HC$_{2\rm n}$CN, the peak abundances during the warming up periods are larger than for the cold period.  On the other hand, for other classes of species (e.g. molecular ions) the issue is less clear.  

\subsection{Chemical Interpretation}\label{sec-cinterp}
\begin{figure}
\includegraphics[angle=90,scale=.8]{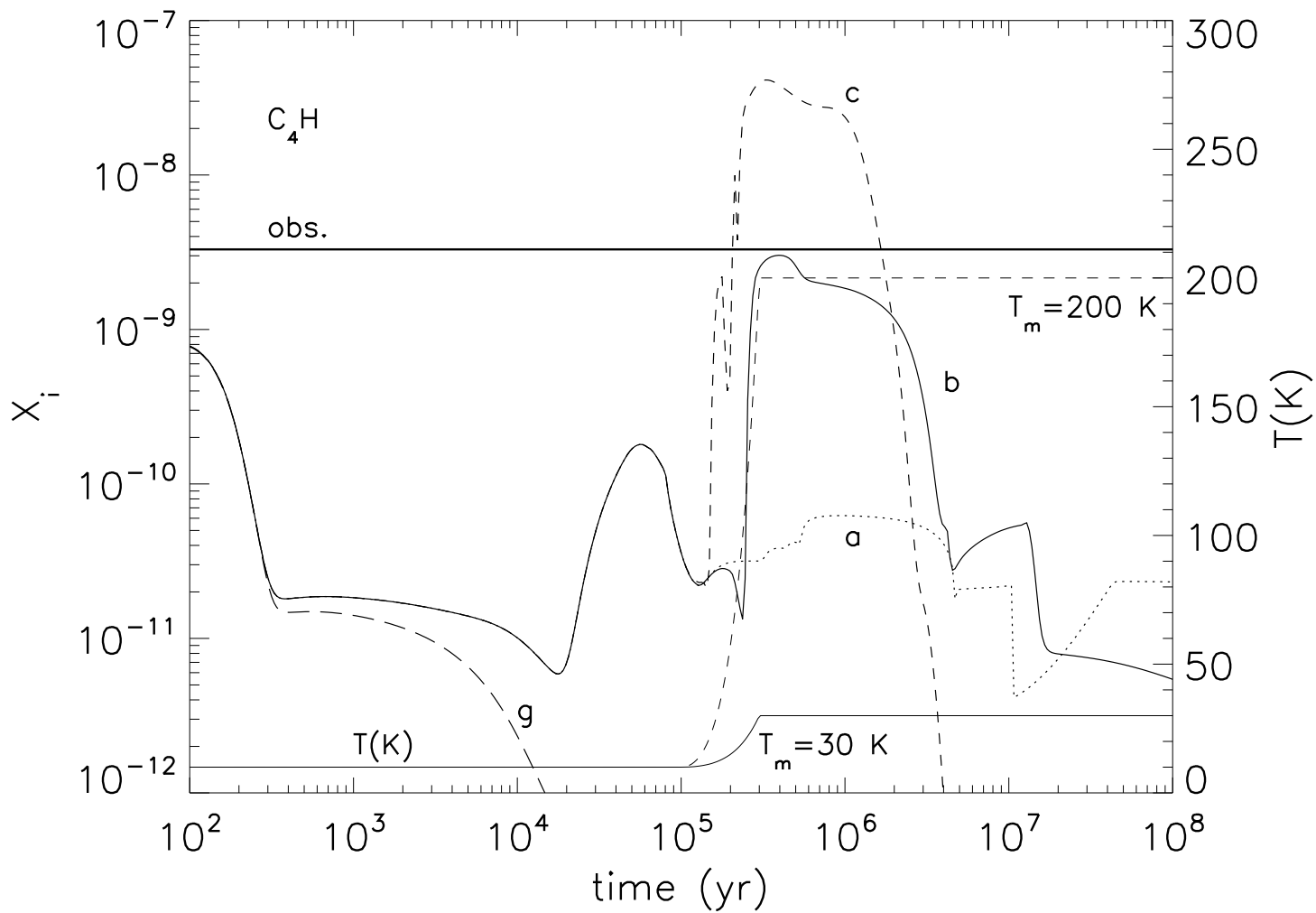}
\caption{Temporal evolution of the $\rm{C_{4}H}$~abundance.  All models undergo a cold pre-stellar core period of $10^{5}$~yr.  The results are computed for three temperature variations: constant $T=10$~K, labeled (a); warm-up to $T_{\rm{max}}=30$~K, labeled (b); warm-up to $T_{\rm{max}}=200$~K, labeled (c).  Models without grain surface chemistry are included (g).  These models overlap for all three heating variations.  The temperature profile for the two warm-up models is also plotted vs time.  The observed abundance is indicated with a horizontal line. \label{fig-c4hex}}
\end{figure}

It is desirable to understand the mechanism(s) responsible for the increase in abundance of gas-phase carbon-chain species during the warm-up period.  As gas-phase species freeze out onto grains in the cold era,  they are regenerated when highly volatile precursor species sublimate as the temperature approaches $\approx 30$~K and subsequently react in the gas phase.  Methane plays a key role in the formation of unsaturated hydrocarbons and so its evolution is shown in Figure~\ref{fig-meth} for models with no warm-up and warm-up to 30 K and 200 K.   During the cold era, a large abundance of methane ice builds up in all models, formed mainly by sequential hydrogenation of carbon on surfaces, in which the concluding reaction is
\begin{equation}
{\rm H(s) + \methrad(s) \longrightarrow \methane(s).}
\end{equation}
 The desorption of methane that occurs as a result of heating to 30 K and beyond becomes apparent in the upper panels of Figure~\ref{fig-meth}. As this desorption occurs, gas-phase chemistry begins to regenerate many unsaturated species.   The corresponding increase in $\rm{C_{4}H}$~abundance is evident in Figure~\ref{fig-c4hex} at $t \sim 3 \times 10^{5}$~yr for $T_{\rm{max}}=30$~K and $t \sim 1.5 \times 10^{5}$~yr for $T_{\rm{max}}=200$~K.  Models without  grain-surface reactions form much less methane (see the abundances of CH$_{4}$(g.o.) in Figure~\ref{fig-meth}), and thus are incapable of regenerating secondary products, as can be seen for C$_{4}$H and HC$_{5}$N in Figures~\ref{fig-c4hex} and~\ref{fig-hc5nex} in the curves labeled ``g''.

The regeneration of unsaturated carbon-chain molecules does not occur as simple sublimation from grain surfaces, but rather as the result of subsequent gas-phase chemistry, similar to that discussed in a dynamic model of the TMC-1 sub-cores by \citet{mmc00}.  For the models with $T_{\rm{max}}=30$~K the most fundamental steps following the sublimation of methane are those leading to the formation of acetylene and related hydrocarbon ions:
\begin{equation}
\rm{C^+} + \methane  \longrightarrow \rm{C_{2}H_{3}^{+}} + \rm{H}, 
\end{equation}  
\begin{equation}
\rm{C^+} + \methane  \longrightarrow \rm{C_{2}H_{2}^{+}} + \rm{H_{2}}, 
\end{equation}
followed by 
\begin{equation}
\rm{C_{2}H_{2}^{+}}  + \rm{H_2}  \longrightarrow \rm{C_{2}H_{4}^{+}}, 
\end{equation}
which lead to the dissociative electron recombination reactions
\begin{equation}
\rm{C_{2}H_{3}^{+}}  + \rm{e^{-}}  \longrightarrow \rm{C_{2}H_{2}} + \rm{H}
\end{equation}
and
\begin{equation}
\rm{C_{2}H_{4}^{+}}  + \rm{e^{-}}  \longrightarrow \rm{C_{2}H_{2}}+ 2\rm{H} . 
\end{equation}
Further reactions among the two-carbon ions and acetylene produce three- and four-carbon atom species.  These products react in turn to form the longer chain $\rm{C_{n}H}$~and $\rm{C_{n}H_2}$~species.   A by-product of the ion molecule network, $\rm{CN}$ will also react with acetylene: 
\begin{equation}
\rm{CN} + \rm{C_{2}H_{2}}  \longrightarrow \rm{HC_{3}N} + \rm{H}, 
\end{equation} 
which is the primary formation path of the first member of the polyyne family.  Further, more complex reaction pathways form longer $\rm{C_{n}N}$~and $\rm{HC_{2n}CN}$~species.     

In models where the temperature increases to 100-200 K, the unsaturated chemistry discussed here is combined with the hot-core chemistry discussed in \citet{gh06}, in which a radical-based surface chemistry followed by the desorption of less vaporizable molecules helps to produce the oxygen-rich organic molecules detected in hot cores.  The observed and peak calculated abundances of some typical hot core molecules such as $\methanol$, $\formaldehyde$, and~$\vinylcyanide$~are included in Table~\ref{tbl-hotcore}, where it can be seen that the lukewarm heating is insufficient to reproduce the observations of typical hot cores. 

Our warm-up models treat the desorption of surface species as if they were pure ices.  In reality, mixtures of ices act in a more complex manner.
  The surface experiments of \citet{cacdvwm04}~address the trapping of more volatile species  where they are either deposited as a surface layer on top of amorphous water ice, or co-deposited with water.  The results indicate that a significant fraction of the most volatile molecules, CO, $\rm{N_{2}}$, and $\rm{O_{2}}$,~can be partially retained in the icy mantle to temperatures much greater than their typical sublimation temperatures.    Such an effect has been considered in the hot-core models of \citet{viti04}.  Our gas-grain code currently does not model surfaces in sufficient detail to account for this retentive effect of water; however, comparison with the other molecules would indicate that not more than half of the available $\methane$(s) would likely be retained.  The fraction released to the gas phase under these conditions still represents a tremendous enhancement over the amount of $\methane$(g) formed by gas phase processes only.  Moreover, a gas-phase model in preparation by Harada \& Herbst shows that the production of unsaturated species from methane is not strongly affected by such a diminution in the amount of gaseous methane vaporized in the vicinity of 30 K. 

\subsection{Comparison with observations}\label{sec-compobs}  
\begin{figure}
\includegraphics[angle=90,scale=.8]{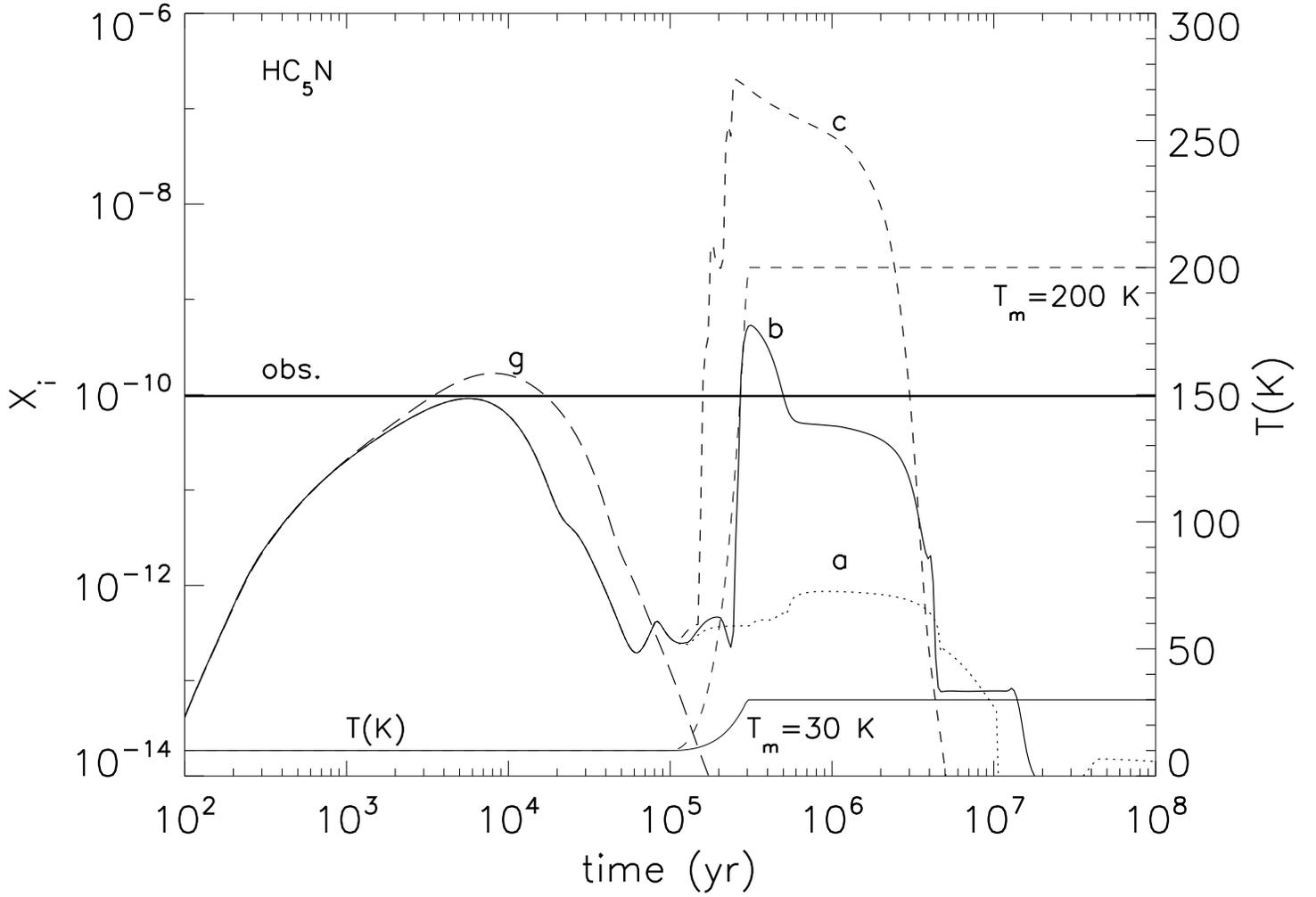}
\caption{Temporal evolution of the $\rm{HC_{5}N}$~abundance.  The labels are the same as those used for Fig.~\ref{fig-c4hex}.\label{fig-hc5nex}}
\end{figure}

\begin{figure}
\includegraphics[angle=90,scale=.8]{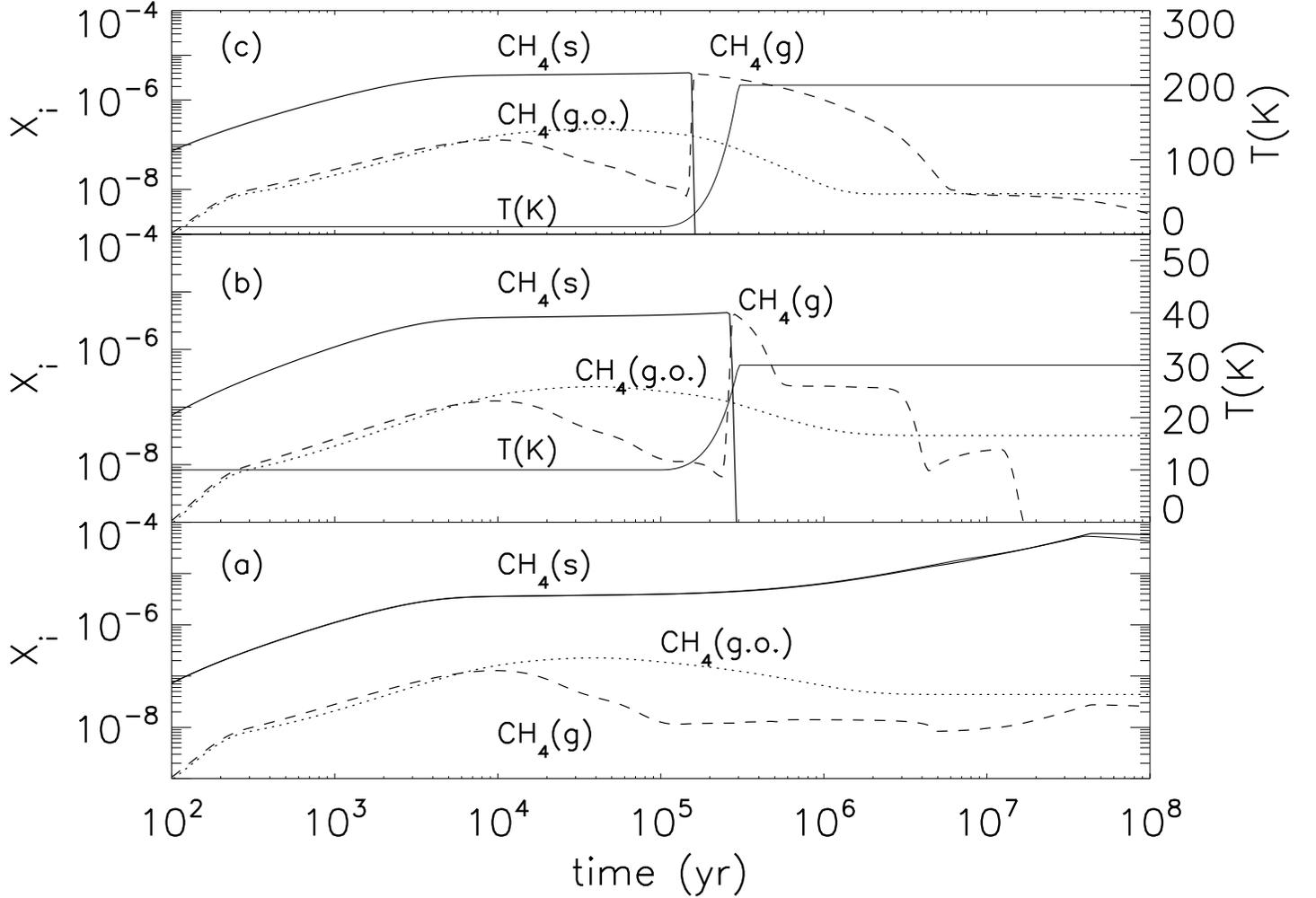}
\caption{Abundance evolution of $\methane$~with heating.  Solid lines represent $\methane$(s) and dashed lines represent $\methane$(g) in gas-grain models. For comparison, dotted lines labeled $\methane$(g.o.) represent $\methane$(g) formed by gas-phase processes only.  Panel (a) contains results for constant $T=10$~K,  panel (b) for warm-up to $T_{\rm{max}}=30$~K, including the temperature profile, and panel (c)  for warm-up to $T_{\rm{max}}=200$~K, also with a temperature profile. \label{fig-meth}}
\end{figure}

Figures \ref{fig-c4hex} and \ref{fig-hc5nex} have an important contrast that helps to clarify the major question of this paper.  The cold ``early-time'' peak of ${\rm{C_{4}H}}$~is less than the observation by more than an order of magnitude, and only the warm-up models produce enough C$_{4}$H to agree with the observation.  However, the observed abundance of ${\rm{HC_{5}N}}$~can be fit to within an order of magnitude with warm-up models or with the cold early-time peak.  We therefore compare the results of all the observed species to determine if the composition of L1527 requires heating to match the observed composition, or if the results are ambivalent.   The reaction network does not differentiate between the normal and carbene isomers of the $\rm{C_{n}H_{2}}$ molecules, which have been detected in L1527.  We assume that the observable vinylidene isomer H$_{2}$C$_{3}$~represents 2\% of the total abundance of $\rm{C_{3}H_{2}}$, as calculated for a dark cloud using the gas-phase network described in \citet{pwh06}.  We extend this estimate to the entire group of $\rm{C_{n}H_{2}}$~species in order to compare the calculations with the abundances of isomers that were actually observed.  

\begin{figure}
\includegraphics[angle=90,scale=.8]{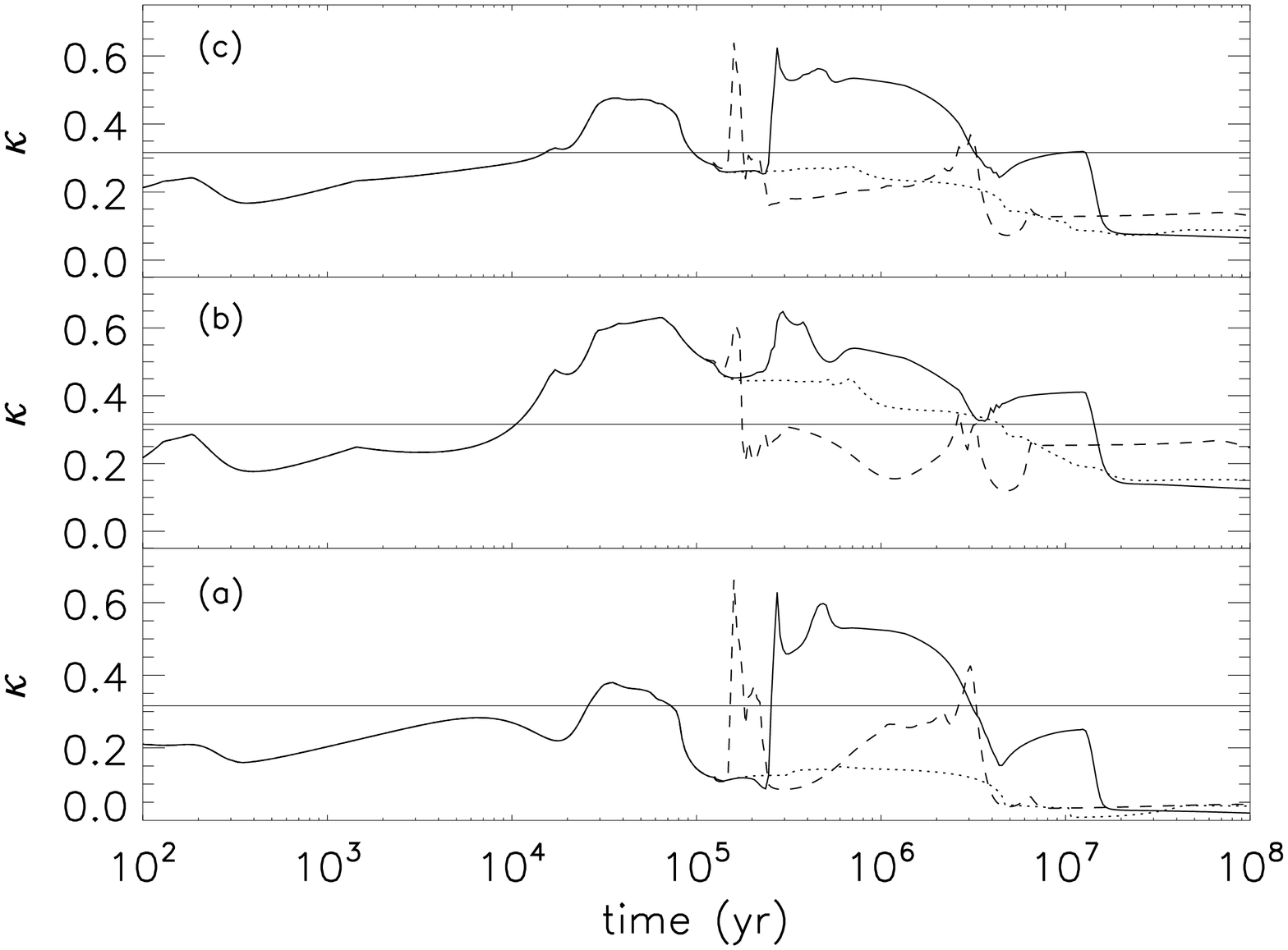}
\caption{Mean confidence level $\kappa$~for observed molecules in L1527 as a function of time.  Results for 13 carbon-chain species observed by  \citet{sea07}~in panel (a), for 9 other species observed by \citet{jsv04}~in panel (b), and for all 22 observed species in panel (c).  The solid line represents  $T_{\rm{max}}=30$~K models,  the dashed line represents $T_{\rm{max}}=200$~K models, and the dotted line represents cold models.  The horizontal line represents one order of magnitude average difference between observed and calculated abundances.\label{fig-conflev}}
\end{figure}

To formulate a quantitative answer to the cold-warm up dichotomy, we revisit the ``mean confidence level'' method utilized by \citet{gwh07}.  The confidence level, $\kappa_{i}$, for agreement between the computed abundance, $X_i$, and the observation, $X_{\rm{obs},i}$, for each relevant molecule is computed as  
\begin{equation}
\kappa_{i}=erfc \left( { | \log (X_{i}) - \log (X_{\rm{obs},i}) | } \over  { {\sqrt{2}}\sigma } \right),
\end{equation} 
where the agreement improves as $\kappa_{i}$~approaches 1.  As before, we define $\sigma=1$~so that $\kappa_{i}=0.317$~indicates a one order of magnitude departure from the observed value.  In addition to the SSHY observations, we include the JSV04 observations to analyze the composition of the cloud.  The  average value $\kappa$ of the confidence level for each set of observations and for all observations combined is shown in Figure~\ref{fig-conflev} as a function of time.  

Within this figure, several meaningful peaks arise.  Prior to the heating, the mean confidence level peaks at early-time for all three data sets in the region of the cold peak.  For the heated models, a peak appears at $t \approx 3\times 10^{5}$~yr for $T_{\rm{max}}=30$~K and at $t \approx 1.5\times 10^{5}$~yr for $T_{\rm{max}}=200$~K.  These peaks indicate that the cold models show an overall agreement within one order of magnitude of the observation, and that the heated models also agree with observation to a similar extent when $T\approx 30$~K.   There is also a secondary peak for the 200 K warm-up models at higher temperatures, which indicates that abundances have increased to overly large values compared with observation as the temperature exceeds 30 K, and then decrease once again.

Now let us compare the data sets more carefully. The bottom panel in Figure \ref{fig-conflev} displays the SSHY observations of unsaturated carbon-chain and cyanopolyyne molecules.  Here it can be seen that the confidence level peaks associated with heating exceed that in the cold period.  Specifically, the confidence level associated with heating is elevated to 0.628 at $t=2.7\times 10^{5}$~yr and $T=25$~K for the $T_{\rm{max}}=30$~K model.  The peak confidence level for the $T_{\rm{max}}=200$~K model is 0.663 at $t=1.6\times 10^{5}$~yr and $T=27$~K, indicating average agreement for heated models to within a factor of 3.  The confidence level of the cold peak is 0.380 at $t=3.5\times 10^{4}$~yr, indicating agreement to within a factor of 7.5.  In this set of observations, heating gives an overall better fit to the observed data.  The middle panel displays the JSV04 observation set.  This plot shows an ambivalence between the cold peak and the heated models, as the maximum confidence levels are 0.631 for the cold peak at $t=6.5\times 10^{4}$~yr, 0.648 for $T_{\rm{max}}=30$~K at $t=2.9\times 10^{5}$~yr and $T=29$~K, and 0.606 for $T_{\rm{max}}=200$~K and $T=27$~K at $t=1.6\times 10^{5}$~yr, indicating agreement to within a factor of 3 for all three possibilities.  The two sets of observations are combined in the top panel, where the warmed models show slightly higher confidence levels of 0.624 and 0.639 as compared with 0.476 (factor of 5) for cold processes.  The optimal agreement peaks for the combined data set occur at the same times as for the SSHY observations.  

This analysis marginally favors the explanation that the observed L1527 composition results at least partially from heating, although it should be noted that the very high confidence level for the warm-up peak lasts only a brief period of time ($\approx 10^{4}$ yr), especially for models with higher maximum temperatures, where the post-peak drop is precipitous.  The significance of dividing this confidence analysis by observation set is that the SSHY results included a profile mapping procedure to demonstrate that the velocity of C$_4$H~observations is associated with the central protostar and spread throughout the gaseous envelope of L1527, rather than associated with the outflows.  The JSV04 observations do not distinguish systematic velocity components.  Future study of this region with detailed interferometry  would certainly add more spatial abundance information and likely better resolve the dichotomy between cold and warm explanations for the composition.

\begin{figure}
\includegraphics[angle=90,scale=.8]{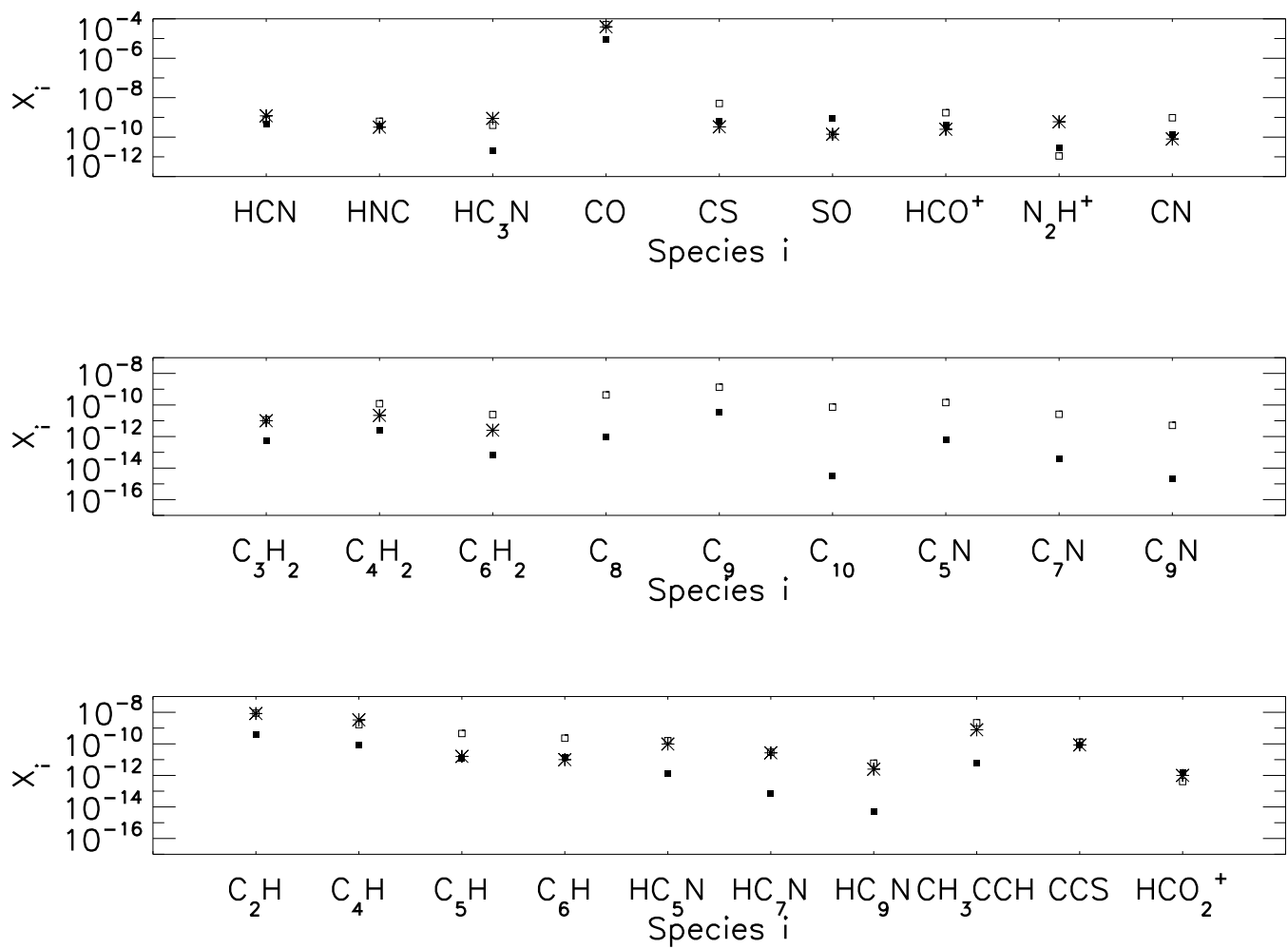}
\caption{Calculated abundances of various carbon-chain and other molecules at the time of optimum agreement between models and observations for warm-up to $T_{\rm{max}}=30$~K.  Abundance results for $T_{\rm{max}}=30$~K ($\square$) at $t=2.7\times 10^{5}$ yr are compared with optimal cold values for constant $T=10$~K ($\blacksquare$) at $t=3.5\times 10^{4}$ yr.  Observations ($\ast$) are denoted where appropriate.\label{fig-30Kres}}
\end{figure}

\begin{figure}
\includegraphics[angle=90,scale=.8]{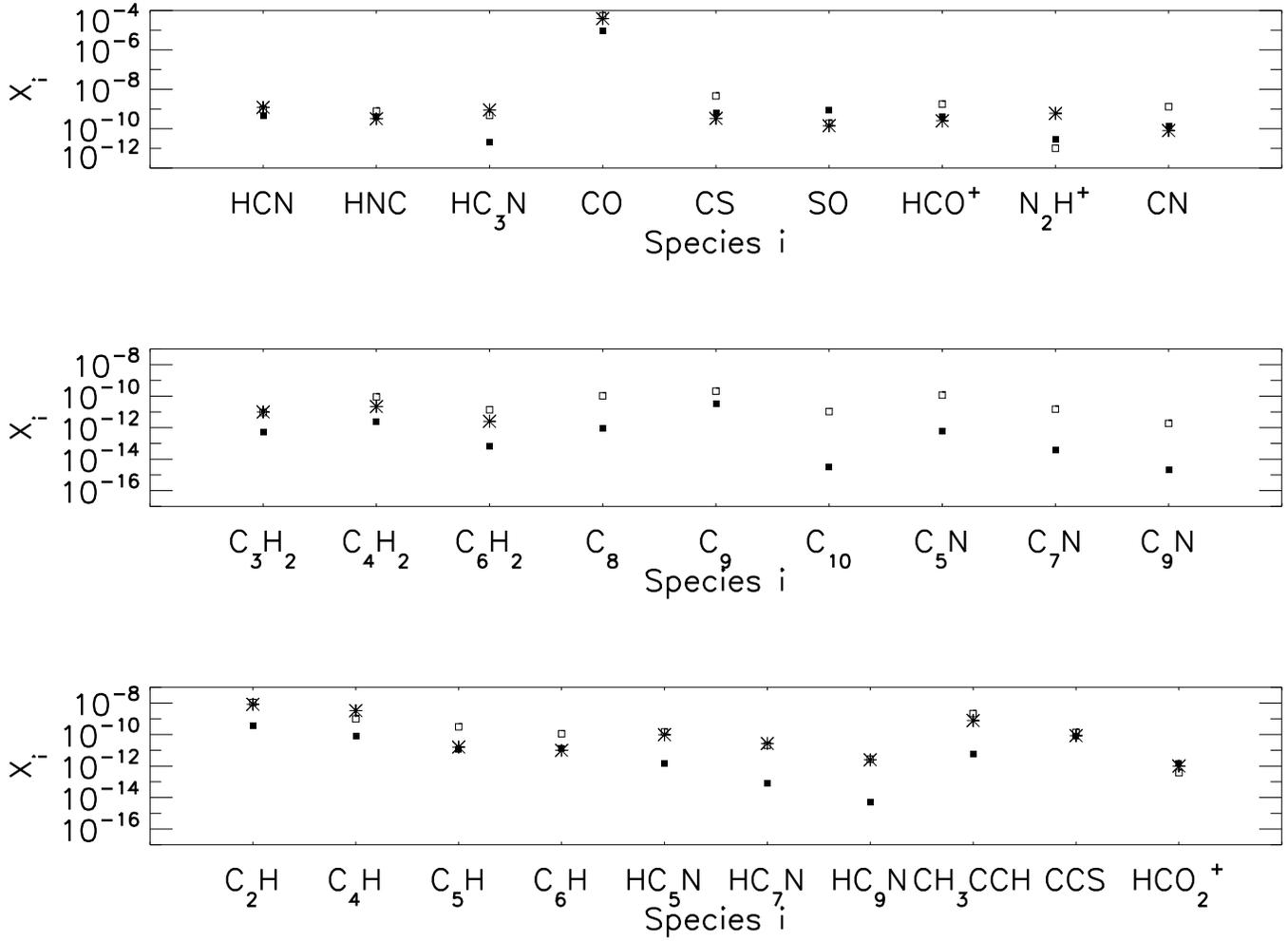}
\caption{Similar results as Fig.~\ref{fig-30Kres}, but for $t=1.6\times 10^{5}$ yr, the time of optimum agreement for warm-up to $T_{\rm{max}}=200$~K.\label{fig-200Kres}}
\end{figure}

Some individual molecular abundances are summarized in Figures {\ref{fig-30Kres}}~and~{\ref{fig-200Kres}}~at the optimum times determined from the above analysis.  Figure {\ref{fig-30Kres}}~compares the optimal warm-up results for $T_{\rm{max}}=30$~K with the optimal cold results, whereas  Figure {\ref{fig-200Kres}}~is a similar comparison for the optimal $T_{\rm{max}}=200$~K results.  The optimal cold results pertain to a cold pre-stellar source with density 10$^{6}$ cm$^{-3}$ and should not be confused with dark cloud abundances such as those of TMC-1.  The bottom panels of these figures display the results for the SSHY species that do not require any adjustment, with the observed values denoted by asterisks where appropriate.  The center panels include the $\rm{C_{n}H_{2}}$ species detected by SSHY and some potentially detectable varieties of carbon-chain molecules.  The top panels display the results relating to observations of JSV04.  In general, these plots also indicate on a visual basis that the SSHY molecules are well reproduced by warm-up models.  Optimal abundances are also listed in Table~\ref{tbl-optimal}, which includes the observed molecules as well as a larger number of predictions.  In this table, italic type indicates that the computed abundance exceeds the observed value by one order of magnitude or more.  Boldface type indicates that the computed abundance is too small by at least one order of magnitude.

From Table~\ref{tbl-optimal}, we can see the individual levels of agreement for observed species at optimum times.  At the optimum time for agreement of warm-up models, a total of 17 of 22 of the observed molecules are within one order of magnitude of the observation, including 11 of the 13 SSHY species and 6 of the 9 from JSV04.  Variations in heating do change the optimum time but not the number of molecules in agreement.  The molecules which do not agree with observation at the optimum time are C$_5$H, C$_6$H, CS, CN, and N$_2$H$^+$, where the models over-produce the first four, but do not make enough N$_2$H$^+$. 

At the optimum time of agreement for the cold-phase model, $t=3.5\times 10^{4}$~yr, only 5 of the 13 SSHY molecules agree with observation to within an order of magnitude, while 8 of 9 JSV04 molecules agree, for a total of 13 out of the 22.  One reason for this comparatively worse agreement can be seen in Figures \ref{fig-c4hex}~and~\ref{fig-hc5nex}, where the cold peaks arise at different times.  Overall, cold processes in this model tend to under-produce the carbon chain molecules observed by SSHY, and cannot produce C$_4$H or HC$_9$N to within an order of magnitude of the observation at any time.  The superiority of the warm-up model for carbon-chain species is seen dramatically in  Figures {\ref{fig-30Kres}}~and~{\ref{fig-200Kres}}.  As the carbon chain molecules get longer, the abundances relative to the cold values are drastically enhanced.  This effect occurs to such an extent that longer chain molecules of the classes $\rm{C_{n}H}$, $\rm{C_{n}H_{2}}$, and $\rm{HC_{2n}CN}$ have only  slighter lower abundances than the shorter chains in that group, at least on the scale depicted here.  The trend presents a good criterion for prediction of further observations of lukewarm corinos.  
As shown in Figure 6 of SSHY, the abundances of carbon-chain species observed in L1527 tend to be slightly lower than those observed in the cold core TMC-1.  These latter observations are fit by cold models in which carbon-rich elemental abundances must be used \citep{shc04,wake06}, or models in which non-thermal mantle evaporation occurs \citep{mmc00}.  

Among the molecules detected in L1527, two stand out as almost uniquely abundant:  C$_{2}$H and C$_{3}$H$_{2}$.  SSHY demonstrated that the abundance of C$_2$H is remarkably high in comparison with several other star-forming and starless cores on the basis of a previous survey \citep{suz92}.  The calculated abundance of C$_2$H  for $T_{\rm{max}}=30$~K remains elevated at this temperature at nearly one order of magnitude greater than the cold peak value for several million years.  However, the abundance of the $T_{\rm{max}}=200$~K model peaks and then rapidly declines to the level of the cold value as $T$~approaches 100 K, prior to $t=2\times 10^{5}$~yr.  This unusual result explains that L1527 at $T\approx 30$~K can reasonably have a C$_2$H abundance that is nearly an order of magnitude greater than both colder starless cores and hotter star forming regions.  This finding is in agreement with the C$_{2}$H image of \citet{bshl08} for IRAS 18089-1732, which shows a hole toward the hot core. 

Also, \citet{bcm98} indicated that the abundance of C$_3$H$_2$~was remarkably greater toward L1527 and the TMC-1 sub-cores than toward any other cold sources included in their survey.  Our calculated lukewarm abundance of C$_3$H$_2$~increases by an order of magnitude over the cold abundance and remains elevated for a large period of time.  The high abundances of both C$_2$H and C$_{3}$H$_{2}$ lend support to the explanation that partial mantle removal in L1527 results in systematic enhancements in carbon-chain abundances, although the case is less obvious for TMC-1 \citep{wake06,mmc00}

 \subsection{Observational Predictions}
The peak abundance values for a variety of carbon-chain and other molecules for $T_{\rm{max}}=30$~K are listed in Table~\ref{tbl-peak} while the abundances at the optimal times for L1527 are listed in Table~\ref{tbl-optimal} for all warm-up models.  This information provides observational predictions for L1527, as well as any other potential lukewarm corino sources.   
 
While our model calculations indicate that warming can successfully regenerate carbon-chain species, they also introduce an unexpected result for hot-core and hot-corino chemistry.  Heating to $T_{\rm{max}}=200$~K increases the peak abundances of these molecules by at least an order of magnitude above the $T_{\rm{max}}=30$~K peak (see Table~\ref{tbl-peak}). Moreover, except for the species C$_{2}$H (see above), these high abundances do not decline rapidly.    Previously, carbon-chain species have not been reported in hot core/corino observations with the exception of $\rm{HC_{3}N}$ \citep{wsw99,gniwb00,b87}, to the best of our knowledge.  It is not clear if unsaturated chain hydrocarbons are simply not present in the observed hot cores, or if they are not detected in the complex spectra of these regions, which are typically dominated by so-called ``weeds''.  However, in a recent survey of the hot corino source IRAS 16293+2422, a number of small carbon-chain species have been detected and are likely to exist in the warm region of the source (E. Caux, C. Ceccarelli,  private communications).  This survey lends a degree of support toward our model result of enhanced carbon-chain abundance in hot regions.  However, it is also possible that the high abundances we predict  in hot cores may be the result of a lack of  destruction mechanisms that become important at higher temperatures.

\subsection{Increased Sulfur}

The gas-phase elemental abundance of sulfur in cold dense sources is highly uncertain because there is no evidence for depletion in diffuse clouds, yet use of a high value for cold dense sources tends to worsen agreement with observation \citep{quan08}.  If the sulfur abundance in our model is raised to 1 $\times$ 10$^{-5}$ (see Table~\ref{tbl-ab}), certain results are affected much more strongly than others.  For the cold phase, the major changes are large diminutions in the peak abundances of the C$_{\rm n}$H species and an increase of at least an order of magnitude for the sulfur-bearing species such as CCS.  For the warm-up phase of the $T_{\rm{max}}=30$~K model to lukewarm corino conditions, the results change little, with peak abundances for all species listed in our tables altered at most by a factor of three.  For the $T_{\rm{max}}=100$~K and $T_{\rm{max}}=200$~K models, the peak abundances tend to be reduced for the carbon-chain species, and their post-peak declines tend to occur more quickly, at a time scale of at most 10$^{6}$ yr, reducing the possible discrepancy with observation for carbon-chain species in hot cores of this age. 

With the mean confidence level approach, we see that the optimum $\kappa$ for the observed species in the cold era decreases to 0.350, barely corresponding to order-of-magnitude agreement.  The optimum $\kappa$ for the warm-up models does not change significantly, on the other hand, indicating a stronger preference for lukewarm processes than in the low-sulfur case.

\section{Conclusions}
Using our gas-grain chemical model, we have attempted to distinguish the chemical composition of L1527 from more familiar evolutionary phases.  More specifically, this composition poses the dilemma that the observed gas-phase abundances might have required heating, might be explained by cold processes alone, or might require a combination of the two depending upon exactly where the molecules are located within the source.  We have approached this question by computing a series of one-point model simulations of a dense core that include various degrees of heating after a preliminary cold phase lasting 10$^{5}$ yr.   A mean confidence level analysis indicates that the composition of the cloud is somewhat more likely to be a result of heating and regeneration than of cold processes.  Dividing the observed molecules into two classes -- the carbon-chain species detected by SSHY and the species detected by JSV04 -- we find that the preference for a warm-up model is stronger for the former species although the level of high agreement during the warm-up period is brief.  This preference is strengthened by the use of a high value for the sulfur elemental abundance.

 The best agreement between model and observation occurs as the most volatile molecules are being desorbed from the grain surfaces.  This agreement occurs at $t=2.7\times 10^{5}$~yr and $T=25$~K for the $T_{\rm{max}}=30$~K model and $t=1.6\times 10^{5}$~yr and $T=27$~K for the $T_{\rm{max}}=200$~K model.  Of the volatile species, the most important chemically is $\methane$ because it acts as a precursor for a gas-phase chemistry leading to carbon-chain species, as discussed by SSHY.  The distinctly large abundances of C$_2$H (SSHY) and C$_3$H$_2$ \citep{bcm98} as compared with various other sources provide some additional support for this lukewarm explanation of the L1527 composition.  Further, the gradual drop off in  abundance of  carbon-chain molecules in L1527  with increasing size is fit better with our warm-up than our cold model.  It should be noted, however, that our model here uses oxygen-rich elemental abundances, and that such models cannot reproduce the abundances of complex molecules seen in TMC-1, so may underestimate the extent of molecular complexity under cold conditions in L1527 \citep{wake06}.  Although the envelope of L1527 at 30 K is  warmer than a cold core at 10 K, it is not nearly as hot as a hot core or hot corino.  The chemistry is clearly distinguishable from the more evolved phase, because the  temperature is not high enough to release some products of grain-surface chemistry such as $\methanol$.   If L1527 is typical of a borderline Class 0 / I low mass protostar, then the ``lukewarm'' corino phase could be fairly common.  Observations of L483 show similar evolutionary characteristics including an enhanced dust temperature of $T_{\rm d}=40$~K \citep{tmmb00}; thus, this source may be a good candidate for a molecular survey of warm carbon chain chemistry.

One complication that arises is that the models in which the temperature rises to 100 K or 200 K produce large abundances of most  gaseous carbon-chain species in addition to the standard hot-core species and these abundances linger on timescales of $2-5 \times 10^{6}$~yr after enhancement, unless a high elemental abundance for sulfur is utilized.  The most salient exception to this rule is C$_{2}$H, for which we predict a low abundance in hot cores and corinos. The problem of high carbon-chain abundances in hot cores also occurs in the hydrodynamic model of \citet{aikawa}, in which large abundances of these species are maintained as the gas collapses to the central star and warms up.  The peak results from Table~\ref{tbl-peak}~show that these abundances are somewhat lower than abundances for previously reported hot core molecules (see Table~\ref{tbl-hotcore}), but may be large enough to be detectable unless their widely-spaced spectral lines are hidden by those of weeds.  Unfortunately,  only detections of HC$_3$N toward hot cores have previously been reported in the literature \citep{wsw99,b87,gniwb00}.  Some of these detections are sufficiently strong that one wonders why more complex cyanopolyynes have not yet been detected. Low upper limits towards Orion have been reported to us by several investigators but this source is complex and may not be representative.   A report of a recent survey in progress (E. Caux, private communication) indicates, on the other hand,  that some smaller carbon-chain species are indeed observable in the hot corino IRAS 16293+2422, most likely associated with the warm region of the source.  Further observations of carbon-chain molecules, particularly with lengths of four or more carbon atoms, toward additional hot core or corino sources might strengthen our characterization of L1527.

E.H. would like to thank the National Science Foundation for supporting his research program in astrochemistry and NASA for support to study interstellar ice formation.  R.T.G. thanks the Alexander von
Humboldt Foundation for a research fellowship.


\clearpage

\begin{deluxetable}{lccl}
\tabletypesize{\scriptsize}
\tablecaption{Granular Properties\label{tbl-gr}}
\tablewidth{0pt}
\tablehead{\colhead{Parameter} & & \colhead{value} &} 
\startdata Sticking coefficient & $S$ & 0.5 & \\
Radius & $a$ & 0.1 & $\mu$m \\
Density& $\rho$ & 3 & g cm$^{-3}$ \\
Mass Fraction & $X_{\rm{d}}$ & 0.01 &  \\
Reactive Desorption Efficiency & $a_{\rm{RRK}}$ & 0.01\\ 
\enddata
\end{deluxetable}

\clearpage

\begin{deluxetable}{lc}
\tabletypesize{\scriptsize}
\tablecaption{Initial Abundances\label{tbl-ab}}
\tablewidth{0pt}
\tablehead{\colhead{Species $i$} & \colhead{${n_i}/{n_{\rm{H}}}$}\tablenotemark{\dagger}}
\startdata $\molh$ & 0.5 \\
He & 0.14\\
C$^+$ & 7.3(-5)\\
N & 2.14(-5)\\
O & 1.76(-4)\\
S$^+$ & 8.0(-8)\tablenotemark{a}\\
Na$^+$ & 2.0(-9)\\
Mg$^+$ & 7.0(-9)\\
Si$^+$ & 8.0(-9)\\
P$^+$ & 3.0(-9)\\
Cl$^+$  & 4.0(-9)\\
Fe$^+$ & 3.0(-9)\\
\enddata
\tablenotetext{\dagger}{$a(b)=a \times 10^{b}$}
\tablenotetext{a}{We also consider the effect of S$^+$=1.0(-5)}
\end{deluxetable}


\begin{deluxetable}{lcccccc}
\tabletypesize{\scriptsize}
\tablecaption{L1527  \& Peak Calculated Fractional Abundances of Assorted Species \label{tbl-peak}}
\tablewidth{0pt}
\tablehead{\colhead{Species }  & Obs. & \colhead{$T=10$~K} & \colhead{$T_{\rm{max}}=30$~K} & \colhead{$T_{\rm{max}}=100$~K} &\colhead{$T_{\rm{max}}=200$~K}&\colhead{Obs. Ref.}}
\startdata $\rm{C_{2}H}$ & 8.3(-09)& 2.0(-09)&8.9(-09)&1.0(-08)&1.1(-08)& a\\
$\rm{C_{3}H}$ &          & 5.0(-10) & 1.9(-09) & 1.2(-08) & 1.5(-08)\\
$\rm{C_{4}H}$ & 3.3(-09) & 1.8(-10) & 3.0(-09) & 3.3(-08) & 4.1(-08) & a\\
$\rm{C_{5}H}$ & 1.6(-11) & 1.2(-11) & 8.7(-10) & 8.1(-09) & 2.9(-08) & a\\
$\rm{C_{6}H}$ & 1.0(-11) & 1.5(-11) & 6.0(-10) & 5.8(-09) & 3.5(-08) & a\\
$\rm{C_{7}H}$ &          & 9.3(-13) & 4.4(-10) & 2.7(-09) & 3.6(-08)\\
$\rm{C_{8}H}$ &          & 1.2(-12) & 7.2(-10) & 3.1(-09) & 3.1(-08)\\
$\rm{C_{9}H}$ &          & 4.8(-14) & 5.1(-10) & 2.4(-09) & 3.4(-08)\\
\hline
$\rm{HC_{5}N}$ & 9.7(-11) & 9.1(-11) & 5.4(-10)& 5.1(-08)& 2.0(-07) & a\\
$\rm{HC_{7}N}$ & 2.7(-11) & 4.3(-12) & 1.9(-10)& 6.3(-09)& 4.1(-08) & a\\
$\rm{HC_{9}N}$ & 2.5(-12) & 1.5(-13) & 9.9(-11)& 1.7(-09)& 1.7(-08) & a\\
\hline
$\rm{C_{2}H_{2}}$ &          & 6.5(-09) & 4.2(-08) & 1.6(-07) & 2.3(-07)\\
$\rm{C_{3}H_{2}}$ & 1.0(-11) & 9.5(-13) & 1.8(-11) & 4.8(-10) & 4.7(-10) & a,b\\
$\rm{C_{4}H_{2}}$ & 2.2(-11) & 2.9(-12) & 1.7(-10) & 1.4(-09) & 2.2(-09)  & a,b\\
$\rm{C_{5}H_{2}}$ &          & 1.4(-13) & 4.8(-11) & 1.3(-09) & 1.4(-09)  & b\\
$\rm{C_{6}H_{2}}$ & 2.5(-12) & 1.2(-13) & 6.2(-11) & 2.0(-10) & 1.0(-09)  & a,b\\
$\rm{C_{7}H_{2}}$ &          & 9.9(-15) & 2.1(-11) & 1.4(-10) & 6.0(-10) & b \\
$\rm{C_{8}H_{2}}$ &          & 8.7(-15) & 3.0(-11) & 7.1(-11) & 4.0(-10) & b \\
$\rm{C_{9}H_{2}}$ &          & 5.9(-16) & 2.0(-11) & 9.7(-11) & 1.1(-09)  & b \\ 
\hline
$\rm{CH_{3}CCH}$    & 7.8(-10) & 2.4(-10) & 4.0(-09) & 9.6(-07) & 9.3(-07)  & a\\
$\rm{CH_{3}C_{4}H}$ &          & 2.0(-12) & 2.3(-09) & 7.2(-09) & 5.9(-08)\\
$\rm{CH_{3}C_{6}H}$ &          & 3.1(-13) & 5.8(-10) & 2.1(-09) & 7.6(-09)\\
$\rm{CCS}$          & 8.5(-11) & 2.0(-10) & 1.3(-10) & 1.2(-09) & 1.1(-09)  & a\\
$\rm{HCO_{2}^{+}}$  & 1.0(-12) & 1.6(-12) & 7.8(-13) & 2.4(-11) & 3.8(-12) & c\\
\hline
$\rm{HCN}$        &  1.2(-09) & 2.1(-07)  & 6.9(-09)  & 1.7(-06)  & 1.8(-06)  & d\\
$\rm{HNC}$        &  3.2(-10) & 1.4(-07)  & 5.6(-09)  & 7.7(-07)  & 1.2(-06)  & d \\
$\rm{HC_{3}N}$    &  8.9(-10) & 3.1(-09)  & 8.8(-10)  & 1.2(-07)  & 1.4(-07)  & d\\
$\rm{CO}$         &  3.9(-05) & 3.3(-05)  & 5.2(-05)  & 7.3(-05)  & 7.3(-05)  & d\\
$\rm{CS}$         &  3.3(-10) & 6.4(-08)  & 8.0(-09)  & 5.7(-08)  & 5.0(-08)  & d\\
$\rm{SO}$         &  1.4(-10) & 1.1(-09)  & 1.5(-10)  & 5.6(-08)  & 5.0(-08)  & d\\
$\rm{HCO^{+}}$    &  6.0(-10) & 4.4(-10)  & 1.8(-09)  & 2.2(-09)  & 1.8(-09)  & d\\
$\rm{N_{2}H^{+}}$ &  2.5(-10) & 6.3(-11)  & 6.1(-10)  & 6.2(-11)  & 8.1(-11) & d\\
$\rm{CN}$         &  8.0(-11) & 3.2(-09)  & 4.0(-09)  & 3.0(-07)  & 3.0(-07)  & d\\
\hline
$\rm{C_{2}}$  & & 3.4(-09)  & 2.0(-09)  & 4.3(-08) & 3.6(-08)\\
$\rm{C_{3}}$  & & 2.3(-06)  & 3.9(-08)  & 2.7(-07) & 2.0(-07)\\
$\rm{C_{4}}$  & & 8.2(-09)  & 1.3(-09)  & 1.2(-08) & 1.8(-08)\\
$\rm{C_{5}}$  & & 1.4(-08)  & 1.5(-08)  & 6.2(-08) & 2.4(-07)\\
$\rm{C_{6}}$  & & 6.0(-11)  & 3.3(-09)  & 5.3(-09) & 5.8(-08)\\
$\rm{C_{7}}$  & & 6.4(-10)  & 7.8(-09)  & 6.9(-09) & 1.7(-07)\\
$\rm{C_{8}}$  & & 5.8(-12)  & 2.9(-09)  & 2.3(-09) & 5.1(-08)\\
$\rm{C_{9}}$  & & 4.4(-11)  & 1.4(-08)  & 7.4(-09) & 2.7(-07)\\
$\rm{C_{10}}$ & & 8.2(-14)  & 9.4(-10)  & 1.4(-09) & 1.5(-08)\\
\hline
$\rm{C_{2}N}$ & & 1.2(-10) & 8.9(-10) & 3.3(-08)  & 4.5(-08)\\
$\rm{C_{3}N}$ & & 2.2(-11) & 3.0(-10) & 3.4(-08)  & 3.3(-08)\\
$\rm{C_{4}N}$ & & 1.6(-11) & 9.2(-10) & 3.2(-08)  & 7.2(-08)\\
$\rm{C_{5}N}$ & & 6.8(-13) & 5.6(-10) & 2.7(-08)  & 6.6(-08)\\
$\rm{C_{7}N}$ & & 4.5(-14) & 2.2(-10) & 3.0(-09)  & 2.0(-08)\\
$\rm{C_{9}N}$ & & 2.8(-15) & 1.1(-10) & 4.8(-10)  & 7.3(-09)\\
\hline
$\rm{CH_{3}C_{3}N}$ & & 7.5(-12) & 7.3(-13) & 1.9(-11) & 1.8(-10)\\
$\rm{CH_{3}C_{5}N}$ & & 2.1(-13) & 8.8(-14) & 2.1(-11) & 2.7(-10)\\
$\rm{CH_{3}C_{7}N}$ & & 9.5(-15) & 3.2(-14) & 2.7(-12) & 3.2(-11)\\
\hline
$\rm{N_{2}}$      & & 1.9(-06)  & 5.6(-06)  & 1.1(-05) & 1.1(-05)\\
$\rm{O_{2}}$      & & 3.7(-07)  & 3.7(-08)  & 4.8(-05) & 4.3(-05)\\
$\rm{CO_{2}}$     & & 5.4(-07)  & 8.6(-09)  & 9.2(-06) & 1.8(-06)\\
$\rm{C_{3}H_{3}}$ & & 2.1(-10)  & 3.9(-09)  & 7.5(-08) & 8.5(-08)\\
$\rm{C_{4}H_{4}}$ & & 3.7(-11)  & 6.1(-10)  & 2.6(-07) & 2.3(-07)\\
$\rm{C_{6}H_{6}}$ & & 5.4(-13)  & 1.6(-09)  & 6.6(-09) & 1.8(-08)\\
$\rm{H_{2}CS}$    & & 1.9(-10)  & 5.7(-10)  & 4.8(-08) & 4.6(-08)\\
$\rm{H_{2}S}$     & & 1.5(-09)  & 4.7(-09)  & 8.7(-09) & 7.9(-09)\\
$\rm{CH_{2}CN}$   & & 1.9(-10)  & 1.0(-10)  & 6.4(-08) & 8.4(-08)\\
$\rm{CH_{2}CO}$   & & 6.4(-09)  & 1.0(-09)  & 1.3(-07) & 1.4(-07)\\
\enddata
\tablenotetext{a}{reported in \citet{sea07}}
\tablenotetext{b}{calculations adjusted for linear isomer as described in \S~\ref{sec-compobs}}
\tablenotetext{c}{unpublished, recently observed by N. Sakai (private communication)}
\tablenotetext{d}{reported in \citet{jsv04}}
\end{deluxetable}

\begin{deluxetable}{lcccccc}
\tabletypesize{\scriptsize}
\tablecaption{Observed  \& Peak Calculated Fractional Abundances of Assorted Hot Core/Corino Species \label{tbl-hotcore}}
\tablewidth{0pt}
\tablehead{\colhead{Species }  & Obs. & \colhead{$T=10$~K} & \colhead{$T_{\rm{max}}=30$~K} & \colhead{$T_{\rm{max}}=100$~K} &\colhead{$T_{\rm{max}}=200$~K}&\colhead{Source}}
\startdata 
$\rm{HCN}$      &  1(-06)$^{a}$ & 2.1(-07) & 6.9(-09)& 1.7(-06)& 1.8(-06) &G327.3-0.6 \\
$\rm{CH_{3}CN}$ &  7(-07)$^{a}$ & 3.4(-09) & 1.5(-11)& 7.4(-08)& 8.3(-08) &G327.3-0.6\\
& [1.6$\pm$0.2](-09)$^{b}$& & & & & NGC 1333 IRAS 4A$^{e}$\\
& [1.0$\pm$0.4](-09)$^{c}$& & & & & IRAS 16293 2422$^{e}$ \\
$\rm{CH_{3}OH}$ & 2(-05)$^{a}$& 6.0(-10) & 6.9(-11)& 2.5(-07)& 1.2(-06)&G327.3-06\\
& $\leq 7(-09)$$^{b}$& & & & & NGC 1333 IRAS 4A$^{e}$\\ 
& 3(-07)$^{c}$& & & & & IRAS 16293 2422$^{e}$\\$
\rm{CH_{3}OCH_{3}}$ & $\leq 2.8(-08)$$^{b}$& 7.7(-15) & 3.8(-16)& 4.2(-10)& 3.2(-09)&NGC 1333 IRAS 4A$^{e}$\\ 
& [2.4$\pm$3.7](-07)$^{c}$& & & & & IRAS 16293 2422$^{e}$ \\
$\rm{H_{2}CO}$ & 2(-08)$^{d}$& 1.2(-09) & 4.3(-09)& 3.8(-08)& 3.7(-08)&NGC 1333 IRAS 4A$^{e}$\\ 
& 1(-07)$^{c}$& & & & & IRAS 16293 2422$^{e}$\\ 
$\rm{HC_{3}N}$ &  3(-11)$^{a}$ & 3.1(-09) & 8.8(-10)& 1.2(-07)& 1.4(-07)&G327.3-0.6\\
&  1.6(-09)$^{f}$ & &       &        &         & Orion KL\\
$\rm{HCOOH}$ & [4.6$\pm$7.9](-09)$^{b}$& 1.8(-10) & 6.0(-12)& 5.1(-09)& 1.6(-08)&NGC 1333 IRAS 4A$^{e}$\\ 
& 6.2(-08)$^{c}$& & & & & IRAS 16293 2422$^{e}$\\
$\rm{HCOOCH_{3}}$ & [3.4$\pm$1.7](-08)$^{b(i)}$& 1.5(-15) & 3.1(-15)& 8.3(-11)& 4.7(-10)&NGC 1333 IRAS 4A$^{e}$\\ 
& [3.6$\pm$0.7](-08)$^{b(ii)}$& & & & & NGC 1333 IRAS 4A$^{e}$\\ 
& [1.7$\pm$0.7](-07)$^{c(i)}$& & & & & IRAS 16293 2422$^{e}$\\
& [2.3$\pm$0.8](-07)$^{c(ii)}$& & & & & IRAS 16293 2422$^{e}$\\ 
$\rm{HNCO}$ & 3(-09)$^{a}$& 4.6(-09) & 4.5(-12)& 7.6(-07)& 8.3(-07)&G327.3-0.6\\ 
\enddata
\tablenotetext{a}{\citet{gniwb00}}
\tablenotetext{b}{ \citet{bclwcccmpt04}: (i) A substate of $\rm{HCOOCH_{3}}$~(ii) E substate of $\rm{HCOOCH_{3}}$}
\tablenotetext{c}{\citet{cea03}: (i) A substate of $\rm{HCOOCH_{3}}$~(ii) E substate of $\rm{HCOOCH_{3}}$}
\tablenotetext{d}{\citet{mea04}}
\tablenotetext{e} {hot corino}
\tablenotetext{f}{\citet{b87}}
\end{deluxetable}

\begin{deluxetable}{lccccc}
\tabletypesize{\scriptsize}
\tablecaption{Observed and Optimal Time Fractional Abundances \label{tbl-optimal}}
\tablewidth{0pt}
\tablehead{ & & \colhead{$T_{\rm{max}}=10$~K} &  \colhead{$T_{\rm{max}}=30$ K}  & \colhead{$T_{\rm{max}}=100$ K} & \colhead{$T_{\rm{max}}=200$ K}\\
\colhead{Species}& Obs.\tablenotemark{a} & \colhead{$X_{i,\rm{opt}}$} &  \colhead{$X_{i,\rm{opt}}$} &\colhead{$X_{i,\rm{opt}}$} &\colhead{$X_{i,\rm{opt}}$}}
\startdata $\rm{C_{2}H}$ & 8.3(-09)& \bf 3.9(-10) & 8.9(-09)  & 1.0(-08) & 1.1(-08)\\
$\rm{C_{3}H}$ &          & 4.9(-10) & 1.0(-09)  & 8.6(-10) & 8.4(-10)\\
$\rm{C_{4}H}$ & 3.3(-09)  & \bf 8.2(-11)\tablenotemark{b} & 1.7(-09)  & 1.1(-09)  & 9.9(-10)\\
$\rm{C_{5}H}$ & 1.6(-11) & 1.2(-11) & \it 4.6(-10) \tablenotemark{c}& \it 3.3(-10) & \it 3.1(-10)\\
$\rm{C_{6}H}$ & 1.0(-11) & 1.3(-11) & \it 2.3(-10) & \it 1.3(-10) & \it 1.1(-10)\\
$\rm{C_{7}H}$ &          & 7.7(-13) & 1.3(-10) & 6.9(-11) & 6.1(-11)\\
$\rm{C_{8}H}$ &          & 5.9(-13) & 1.1(-10) & 4.5(-11) & 3.6(-11)\\
$\rm{C_{9}H}$ &          & 4.1(-14) & 7.2(-11) & 2.9(-11) & 2.3(-11)\\
\hline
$\rm{HC_{5}N}$ & 9.7(-11) & \bf 1.4(-12) & 1.6(-10) & 1.4(-10) & 1.5(-10)\\
$\rm{HC_{7}N}$ & 2.7(-11) & \bf 7.5(-14) & 2.9(-11) & 2.1(-11) & 2.0(-11)\\
$\rm{HC_{9}N}$ & 2.5(-12) & \bf 5.0(-15) & 5.9(-12) & 3.0(-12) & 2.7(-12)\\
\hline
$\rm{C_{2}H_{2}}$ &          & 6.5(-09)  & 3.7(-08)  & 3.3(-08)  & 3.2(-08)\\
$\rm{C_{3}H_{2}}$ & 1.0(-11) & \bf 5.0(-13) & 1.1(-11) & 9.6(-12) & 9.3(-12)\\
$\rm{C_{4}H_{2}}$ & 2.2(-11) & 2.4(-12) & 1.2(-10) & 9.6(-11) & 9.0(-11)\\
$\rm{C_{5}H_{2}}$ &          & 1.3(-13) & 2.5(-11) & 1.8(-11) & 1.6(-11)\\
$\rm{C_{6}H_{2}}$ & 2.5(-12) & \bf 6.3(-14) & 2.5(-11) & 1.5(-11) & 1.4(-11)\\
$\rm{C_{7}H_{2}}$ &          & 9.8(-15) & 7.1(-12) & 4.0(-12) & 3.4(-12)\\
$\rm{C_{8}H_{2}}$ &          & 7.1(-15) & 5.6(-12) & 2.6(-12) & 2.1(-12)\\
$\rm{C_{9}H_{2}}$ &          & 5.5(-16) & 2.6(-12) & 1.0(-12) & 7.8(-13)\\
\hline
$\rm{CH_{3}CCH}$    & 7.8(-10) & \bf 5.9(-12) & 2.2(-09)  & 2.1(-09)  & 2.2(-09)\\
$\rm{CH_{3}C_{4}H}$ &          & 1.7(-12) & 1.1(-09)  & 7.0(-10) & 6.2(-10)\\
$\rm{CH_{3}C_{6}H}$ &          & 3.1(-13) & 1.7(-10) & 8.6(-11) & 6.8(-11)\\
$\rm{CCS}$          & 8.5(-11) & 8.2(-11) & 1.3(-10) & 1.4(-10) & 1.4(-10)\\
$\rm{HCO_{2}^{+}}$  & 1.0(-12)   & 1.4(-12) & 4.1(-13) & 3.8(-13) & 3.7(-13)\\
\hline
$\rm{HCN}$        & 1.2(-09)  & 4.6(-10) & 7.7(-10) & 8.3(-10) & 9.3(-10)\\
$\rm{HNC}$        & 3.2(-10) & 3.9(-10) & 6.3(-10) & 7.0(-10) & 7.8(-10)\\
$\rm{HC_{3}N}$    & 8.9(-10) & \bf 2.0(-11) & 4.0(-10) & 4.3(-10) & 4.7(-10)\\
$\rm{CO}$         & 3.9(-05)  & 8.9(-06)  & 5.1(-05)  & 5.4(-05)  & 5.4(-05)\\
$\rm{CS}$         & 3.3(-10) & 6.4(-10) & \it 5.1(-09)  & \it 5.1(-09)  & \it 4.6(-09)\\
$\rm{SO}$         & 1.4(-10) & 9.0(-10) & 1.3(-10) & 2.1(-10) & 1.9(-10)\\
$\rm{HCO^{+}}$    & 6.0(-10) & 4.1(-10) & 1.7(-09)  & 1.7(-09)  & 1.8(-09)\\
$\rm{N_{2}H^{+}}$ & 2.5(-10) & 2.7(-11) & \bf 1.1(-11) & \bf 1.0(-11) & \bf 1.0(-11)\\
$\rm{CN}$         & 8.0(-11) & 1.4(-10) & \it9.6(-10) & \it1.2(-09)  & \it1.3(-09)\\
\hline
$\rm{C_{2}}$  && 2.0(-10) & 1.1(-09)  & 1.1(-09)  & 1.1(-09)\\
$\rm{C_{3}}$  && 1.5(-07)  & 1.1(-08)  & 7.4(-09)  & 6.1(-09)\\
$\rm{C_{4}}$  && 2.8(-11) & 1.0(-09)  & 8.2(-10) & 6.9(-10)\\
$\rm{C_{5}}$  && 1.3(-08)  & 6.7(-09)  & 3.7(-09)  & 2.7(-09)\\
$\rm{C_{6}}$  && 2.1(-11) & 1.1(-09)  & 5.8(-10) & 4.4(-10)\\
$\rm{C_{7}}$  && 6.0(-10) & 2.7(-09)  & 1.1(-09)  & 7.4(-10)\\
$\rm{C_{8}}$  && 9.7(-13) & 4.3(-10) & 1.5(-10) & 1.1(-10)\\
$\rm{C_{9}}$  && 3.4(-11) & 1.3(-09)  & 3.5(-10) & 2.1(-10)\\
$\rm{C_{10}}$ && 3.2(-15) & 7.3(-11) & 1.8(-11) & 1.1(-11)\\
\hline
$\rm{C_{2}N}$ && 9.9(-11) & 4.3(-10) & 4.0(-10) & 3.9(-10)\\
$\rm{C_{3}N}$ && 1.9(-11) & 1.2(-10) & 1.2(-10) & 1.4(-10)\\
$\rm{C_{4}N}$ && 9.9(-12) & 1.7(-10) & 8.2(-11) & 6.3(-11)\\
$\rm{C_{5}N}$ && 5.9(-13) & 1.4(-10) & 1.2(-10) & 1.2(-10)\\
$\rm{C_{7}N}$ && 3.8(-14) & 2.6(-11) & 1.6(-11) & 1.5(-11)\\
$\rm{C_{9}N}$ && 2.2(-15) & 5.1(-12) & 1.2(-12) & 1.9(-12)\\
\hline
$\rm{CH_{3}C_{3}N}$ && 1.9(-14) & 2.9(-13) & 2.7(-13) & 2.7(-13)\\
$\rm{CH_{3}C_{5}N}$ && 1.7(-16) & 2.3(-14) & 1.9(-14) & 1.8(-14)\\
$\rm{CH_{3}C_{7}N}$ && 8.0(-18) & 3.8(-15) & 2.3(-15) & 2.2(-15)\\
\hline
$\rm{N_{2}}$      && 7.6(-07)  & 5.6(-06)  & 5.7(-06)  & 5.7(-06)\\
$\rm{O_{2}}$      && 3.6(-07)  & 1.1(-08)  & 5.1(-09)  & 4.6(-09)\\
$\rm{CO_{2}}$     && 1.7(-07)  & 4.1(-09)  & 2.8(-09)  & 2.4(-09)\\
$\rm{C_{3}H_{3}}$ && 2.3(-12) & 2.2(-09)  & 2.2(-09)  & 2.3(-09)\\
$\rm{C_{4}H_{4}}$ && 4.1(-12) & 9.5(-11) & 7.8(-11) & 8.5(-11)\\
$\rm{C_{6}H_{6}}$ && 3.0(-13) & 3.2(-10) & 1.8(-10) & 1.7(-10)\\
$\rm{H_{2}CS}$    && 4.4(-11) & 5.2(-10) & 6.6(-10) & 7.3(-10)\\
$\rm{H_{2}S}$     && 1.1(-10) & 3.3(-09)  & 7.8(-09)  & 7.7(-09)\\
$\rm{CH_{2}CN}$   && 2.7(-11) & 7.4(-11) & 8.6(-11) & 8.6(-11)\\
$\rm{CH_{2}CO}$   && 2.7(-10) & 1.0(-09)  & 1.2(-09)  & 1.1(-09)\\
\hline
$\rm{CH_{3}CN}$      && 6.3(-13) & 2.0(-12) & 2.3(-12) & 2.6(-12)\\
$\rm{CH_{3}OH}$      && 2.9(-12) & 5.0(-11) & 3.4(-12) & 1.8(-12)\\
$\rm{CH_{3}OCH_{3}}$ && 1.4(-17) & 3.8(-16) & 1.1(-17) & 6.9(-18)\\
$\rm{H_{2}CO}$       && 5.9(-11) & 4.3(-09)  & 4.3(-09)  & 4.5(-09)\\
$\rm{HCOOH}$         && 6.1(-11) & 3.6(-12) & 1.7(-12) & 1.5(-12)\\
$\rm{HCOOCH_{3}}$    && 2.4(-16) & 3.8(-15) & 5.9(-16) & 4.7(-16)\\
$\rm{HNCO}$          && 2.2(-10) & 2.4(-12) & 4.3(-12) & 4.2(-12)\\
\enddata
\tablenotetext{a}{Observational references are the same as those in Table~\ref{tbl-peak}.}
\tablenotetext{b}{Boldface type indicates that the computed abundance is less than the observed value by one order of magnitude or more.}

\tablenotetext{c}{Italic type indicates that the computed abundance is greater than the observed value by one order of magnitude or more.}

\end{deluxetable}


\begin{thebibliography}{}
\bibitem[Ag\'{u}ndez et al.(2008)]{agund08} Ag\'{u}ndez, M., Cernicharo, J., Gu\'{e}lin, M., Gerin, M., McCarthy, M. C., \& Thaddeus, P. 2008, A\&A, 478, L19
\bibitem[Aikawa et al.(2008)]{aikawa} Aikawa, Y., Wakelam, V., Garrod, R.T., \& Herbst, E. 2008, \apj, 674, 984
\bibitem[Andr\'{e}, Ward-Thompson \& Barsony(2000)]{awb00} Andr\'{e}, P., Ward-Thompson, D., \& Barsony, M. 2000, in Protostars and Planets IV, ed. V. Mannings, A.P. Boss, \& S.S. Russell (Tucson: University of Arizona Press), 59
\bibitem[Benson, Caselli, \& Myers(1998)]{bcm98} Benson, P.J., Caselli, P., \& Myers, P.C. 1998, \apj, 506, 743
\bibitem[Beuther et al.(2008)]{bshl08} Beuther, H., Semenov, D., Henning, Th., \& Linz, H. 2008, \apj, 675, L33
\bibitem[Blake et al.(1987)]{b87} Blake, G. A., Sutton, E. C., Masson, C. R., \& Phillips, T. G. 1987, ApJ, 315, 621
\bibitem[Bottinelli et al.(2004)]{bclwcccmpt04} Bottinelli, S., Ceccarelli, C., Neri, R., Williams, J.P., Caux, E., Cazaux, S., Lefloch, B., Maret, S., \& Tielens, A.G.G.M. 2004, \apj, 617, L69
\bibitem[Cazaux et al.(2003)]{cea03} Cazaux, S., Tielens, A. G. G. M., Ceccarelli, C., Castets, A., Wakelam, V., Caux, E., Parise, B., \& Teyssier, D. 2003, \apj, 593, L51
\bibitem[Ceccarelli et al.(2000)]{ccchlmt00} Ceccarelli, C., Castets, A., Caux, E., Hollenbach, D., Loinard, L., Molinari, S., \& Tielens, A.G.G.M. 2000, \aap, 355, 1129
\bibitem[Collings et al.(2004)]{cacdvwm04} Collings, M.P., Anderson, M.A., Chen, R., Dever, J.W., Viti, S., Williams, D.A., \&  McCoustra, M.R.S. 2004, MNRAS, 354, 1133
\bibitem[Garrod \& Herbst(2006)]{gh06} Garrod, R.T., \& Herbst, E. 2006, \aap, 457, 927
\bibitem[Garrod et al.(2006)]{gpch06} Garrod, R.T., Park, I.H., Caselli, P., \& Herbst, E. 2006, in Faraday Discussions, Chemical Evolution of the Universe, 133, 51
\bibitem[Garrod, Wakelam, \& Herbst(2007)]{gwh07} Garrod, R.T., Wakelam, V., \& Herbst, E. 2007, \aap, 467, 1103
\bibitem[Gibb et al.(2000)]{gniwb00} Gibb, E.L., Nummelin, A., Irvine, W.M., Whittet, D.C.B., \& Bergman, P. 2000 \apj, 545, 309
\bibitem[Gibb et al.(2004)]{gibb04} Gibb, E. L., Whittet, D. C. B., Boogert, A. C. A., \& Tielens, A. G. G. M. 2004, ApJS, 151, 35
\bibitem[Hasegawa, Herbst, \& Leung(1992)]{hhl92} Hasegawa, T.I., Herbst, E., \& Leung, C.M. 1992, \apjs, 82, 167
\bibitem[Hirota, Maezawa, \& Yamamoto(2004)]{hmy04} Hirota, T., Maezawa, H., \& Yamamoto, S. 2004, \apj, 617, 399 
\bibitem[Hogerheijde \& Sandell(2000)]{hs00} Hogerheijde, M.R., \& Sandell, G. 2000, \apj, 534, 880
\bibitem[J{\o}rgensen, Sch\"{o}ier, \& van Dishoeck(2004)]{jsv04} J{\o}rgensen, J.K., Sch\"{o}ier, F.L.  \& van Dishoeck, E.F. 2004, \aap, 416, 603 (JSV04)
\bibitem[J{\o}rgensen, Sch\"{o}ier, \& van Dishoeck(2002)]{jsv02} J{\o}rgensen, J.K., Sch\"{o}ier, F.L.  \& van Dishoeck, E.F. 2002 \aap, 389, 908
\bibitem[Loinard et al.(2002)]{lrdwh02} Loinard, L., Rodr\'{i}guez, L., D'Alessio, P., Wilner, D.J., \& Ho, P.T.P. 2002, \apj, 581, L109
\bibitem[Maret et al.(2004)]{mea04} Maret, S.,  Ceccarelli, C., Caux, E., Tielens, A.G.G.M., J{\o}rgensen, J.K., van Dishoeck, E.F., Bacmann, A., Castets, A., Lefloch, B., Loinard, L., Parise, B., \& Sch\"{o}ier, F.L. 2004, \aap, 416, 577
\bibitem[Markwick, Millar, \& Charnley(2000)]{mmc00} Markwick, A.J., Millar, T.J., \& Charnley, S.B. 2000, \apj, 535, 256 
\bibitem[Myers \& Mardones(1998)]{mm98} Myers, P.C., \& Mardones, D. 1998, in ASP Conf. Ser. 132, Star Formation with the Infrared Space Observatory, ed. J. Yun \& R. Liseau (San Francisco: ASP), 173
\bibitem[\"{O}berg et al.(2008)]{oberg08} \"{O}berg, K.I., Boogert, A.C.A., Pontoppidan, K.M., Blake, G.A., Evans, N.J., Lahuis, F., \& van Dishoeck, E.F. 2008, \apj, accepted (astro-ph/0801.1223)
\bibitem[\"{O}berg et al.(2007)]{ofafsvl07} \"{O}berg, K.I., Fuchs, G.W., Awad, Z., Fraser, H.J., Schlemmer, S., van Dishoeck, E.F., \& Linnartz, H. 2007, \apj, 662, L23
\bibitem[Ohashi et al.(1997)]{ohhm97} Ohashi, N., Hayashi, M., Ho, P.T.P., \& Momose, M. 1997, \apj, 475, 211
\bibitem[Park, Wakelam, \& Herbst(2006)]{pwh06} Park, I.H., Wakelam, V., \& Herbst, E. 2006, \aap, 449, 631
\bibitem[Quan et al.(2008)]{quan08} Quan, D., Herbst, E., Millar, T. J., Hassel, G. E., Ling, S. Y.,  Guo, H., Honvault, P., \& Xie, D. 2008, ApJ, in press
\bibitem[Sakai et al.(2008a)]{sea07} Sakai, N., Sakai, T., Hirota, T. \& Yamamoto, S. 2008a, \apj, 672, 371 (SSHY)
\bibitem[Sakai et al.(2007)]{ssoy07} Sakai, N., Sakai, T., Osamura, Y. \& Yamamoto, S. 2007, \apj, 667, L65
\bibitem[Sakai et al.(2008b)]{ssy07} Sakai, N, Sakai, T., \& Yamamoto, S. 2008b, \apj, 673, L71
\bibitem[Shu(1977)]{shu77} Shu, F.H. 1977, \apj, 214, 488
\bibitem[Smith, Herbst, \& Chang(2004)]{shc04} Smith, I.W.S., Herbst, E., \& Chang, Q. 2004, MNRAS, 350, 323
\bibitem[Suzuki et al.(1992)]{suz92} Suzuki, H., Yamamoto, S., Ohishi, M., Kaifu, N., Ishikawa, S., Hirahara, Y., \& Takano, S. 1992, \apj, 392, 551 
\bibitem[Tafalla et al.(2000)]{tmmb00} Tafalla, M., Myers, P.C.., Mardones, D., \& Bachiller, R. 2000, \aap, 359, 967
\bibitem[Viti et al.(2004)]{viti04} Viti, S., Collings, M.P., Dever, J.D., McCoustra, M.R.S., \& Williams, D.A. 2004, MNRAS, 354, 1141
\bibitem[Wakelam et al.(2006)]{wake06} Wakelam, V., Herbst, E., \& Selsis, F. 2006, A\&A, 451, 551
\bibitem[Wyrowski et al.(1999)]{wsw99} Wyrowski, F., Schilke P., \& Walmsley, C. M. 1999, A\&A, 341, 882
\end{thebibliography}
\end{document}